\documentclass[12pt]{article}
\newcommand{\mathsym}[1]{{}}
\def\id{\protect{{1 \kern-.28em {\rm l}}}}

\def\be{\begin{eqnarray}}
\def\ee{\end{eqnarray}}

\makeatletter
\renewcommand\section{\@startsection {section}{1}{\z@}%
                                   {-3.5ex \@plus -1ex \@minus -.2ex}%
                                   {2.3ex \@plus.2ex}%
                                   {\normalfont\large\bfseries}}
\renewcommand\subsection{\@startsection{subsection}{2}{\z@}%
                                   {-3.25ex\@plus -1ex \@minus -.2ex}%
                                   {1.5ex \@plus .2ex}%
                                   {\normalfont\normalsize\bfseries}}
\def\Tr{{\rm Tr}}

\def \foot {\footnote}
\def \bi{\bibitem}

\def \ha {{1 \over 2}}

\def \ci{\cite}

\def \E {{\mathcal  E}} \def \J {{\mathcal  J}}

\def \d {\del}

\def \L {\Lambda}

\def\z{\zeta}

\def\a{\alpha}

\def\C{{\bf C}}
\def\P{{\bf P}}

\def \del{\partial}
\def \a {\alpha}

\def\g{\gamma}
\def\s{\sigma}
\def\z{\zeta}

\def\ov{\over}

\def\J{{\mathcal J}}

\def\E{{\mathcal E}}

\def\l{\lambda}

\def\foot{\footnote}

\def \ci {\cite}

\def \foot {\footnote}
\def \bi{\bibitem}
\def \ha {{1 \over 2}}

\def \fo { {1\ov 4}}

\def \d {\partial}

\def \el {\ell}
\def \Tr {{\rm Tr}}
\def \P {\Phi}
\def \l  {\lambda}

\def \V {v}

\def \D {\Delta}

\def \m {\mu}

\def \bi{\bibitem}
\def \la {\label}

\def \l {\lambda}
\def\foot{\footnote}

\def \adss {$AdS_5 \times S^5$ }
\newcommand{\rf}[1]{(\ref{#1})}
\def \ov {\over}

\def\cc{\circ}

\def \ha{{1\ov 2}}

\def \no {\nonumber}

\def \J {\mathcal{J}}
\def \del {\partial}

\def \E {{\cal E}}

\def \J {{\cal J}}

 \def \bb {\bar \beta}

\def \bi{\bibitem}
\def \la {\label}

\def \l {\lambda}
\def\foot{\footnote}

\def \sql {{\sqrt \l}}

\def \adss {$AdS_5 \times S^5$}

\def \D {\Delta}

\def \ov {\over}

\def \varpi {{\rm w}}
\def \OO {{\cal O}}

\def \pa{\partial}

\def \te {\theta}

\def \cc {{\rm f}}
\def \d {\delta}

\def \pa {\partial}
\def \C {{\cal C}}
\def \bea {\be}
\def \eea {\ee}
  \def \d {\delta}

\def\Tr{{\rm Tr}}

\def \del {\partial} 
\def \d {\partial}
\def \s {\sigma}

 \def \L {{\cal L}}

 \def \J {{\cal J}}
 
 \def \E {{\cal E}}

\def \d {\del}
\def \bd {\bar \del}

\newcommand{\ft}[2]{{\textstyle\frac{#1}{#2}}}

\def \os  {\OO({\textstyle{ 1\ov \sql}})}
\def \oss  {\OO({\textstyle{ 1\ov (\sql)^2}})}

\def \bd {\bar \del} \def \sql {\sqrt{\lambda}} 
\def \vp {\varphi}

\def \cc {{c }} 
\def \OO {{\cal O}}
\def \te {\textstyle}
\def \fl {\sqrt[4]{\l}}

\def \ha {{{\textstyle{1 \ov2}}}}
\def \fo {{\textstyle{1 \ov4}}}
\def \rx {{\rm x}}
\def \hg {{\hat \g}}

\def \C  {{\rm C}}
\def \hC  {{\rm \hat  C}}
\def \dd  {{\rm d}}
\def \bb {{\rm b}}
\def \dDelta {2}
\def \sql {{\sqrt{\l}}}
\def \fl {\sqrt[4]{\lambda}}

\def \be {
\begin{eqnarray}}
\def \ee {\end{eqnarray} }
\def \an {{\rm an}} \def \nan {{\rm nan}}
 
\def \oblue { } 
\def \ored  { }

\def \l {\lambda}
\def \sql {{\sqrt \lambda}}

\begin{document}

\overfullrule=0pt
\parskip=2pt
\parindent=12pt
\headheight=0in \headsep=0in \topmargin=0in \oddsidemargin=0in

\vspace{ -3cm}
\thispagestyle{empty}
\vspace{-1cm}

\begin{center}
\vspace{1cm}
{\Large\bf  

%``Short'' 
Quantum  strings in $AdS_5\times S^5$ \\
\vspace{0.2cm}
 and AdS/CFT duality\foot{
 Based on a talk given at the international workshop
 ``Crossing the boundaries: Gauge dynamics at strong coupling'',
 honouring the 60th birthday of M.A. Shifman, Minneapolis, 
 May 14-17, 2009;  published  in the proceedings in  Int.J.Mod.Phys. A25,  319-331, 2010} 

%   in $N=4$ SYM   theory 
 
\vspace{1.2cm}

   }

\vspace{.2cm}
 {
A.A. Tseytlin$
$\footnote{Also at Lebedev  Institute, Moscow. tseytlin@imperial.ac.uk }}\\

\vskip 0.6cm

{\em 
Blackett Laboratory, Imperial College,
London SW7 2AZ, U.K.
 }

\vspace{.2cm}

\end{center}

\begin{abstract}
 We review some recent progress in understanding the spectrum of energies/dimensions 
of strings/operators  in  \adss -- planar $\cal N$=4 super Yang-Mills  correspondence. 
We consider  leading strong coupling corrections to the energy 
 of  lightest massive string modes in \adss, which should be dual to members of the 
 Konishi operator multiplet  in the  SYM theory. 
 %AT Through AdS/CFT  
 This determines the  general structure of  strong-coupling 
 expansion of the  anomalous   dimension of the  Konishi operator.  
 We  use  1-loop  results for    semiclassical string  states 
 to extract  information about the leading  coefficients in this
  expansion.

%%%%%%%%%%%%%%%%%%%%%%%%%%%%%%%%%
\end{abstract}

%\newpage
%\setcounter{equation}{0} 
%\setcounter{footnote}{0}
%\setcounter{section}{0}

%\renewcommand{\theequation}{1.\arabic{equation}}
 %\setcounter{equation}{0}

%\setcounter{equation}{0} \setcounter{footnote}{0}
%\setcounter{section}{0}

\def \edd {\end{document}}

%\edd
\def \cc {{c }} 
\def \OO {{\cal O}}
\def \te {\textstyle}
\def \fl {\sqrt[4]{\lambda}}

\def \ha {{{\textstyle{1 \ov2}}}}
\def \fo {{\textstyle{1 \ov4}}}
\def \rx {{\rm x}}
\def \hg {{\hat \g}}

\def \C  {{\rm C}}
\def \hC  {{\rm \hat  C}}
\def \dd  {{\rm d}}
\def \bb {{\rm b}}
\def \dDelta {2}
\def \sql {{\sqrt{\lambda}}}

 \def \an {{\rm an}} \def \nan {{\rm nan}}
 
%%%%%%%%%%%%%%%%%%%%%%%%%%%%%%%%%%%%%%

\section{Introduction }
%%%%%%%%%%%%%%%%%%%%%%%%%%%%%%%%%%%
%refs to short ? Ambjorn   Janik  AF   Kazakov   ?
%To study the spectrum of anomalous dimensions we are to compute 2d dimensions 
%and impose marginality condition. 
%We may use standard NSR approach or l.c. gauge  or GS string approaches
%which all are available   in near-flat limit.
%if we look at small  massive string states  the zero-mode issues should not be relevant 
%and we may  compute their masses  using near flat space expansion.

%%%%%%%%%%%%%%%%%%%%%%%%%%%%%%%%%%%%%%%%%%%%%%%%%%%%%%%%%%%%%%%%%%%%%%%%%%%%
%\be\D= 2 \lambda^{1/4}  + b_0   +  b_1 \lambda^{-1/4} +  b_3 \lambda^{-3/4}+ 
 % ...\ee
  
Two important problems of  modern  theoretical high energy physics are 
to   understand quantum gauge theories at any coupling
(with applications to both perturbative and non-perturbative issues) 
and to understand string theories in non-trivial backgrounds 
(e.g. Ramond-Ramond  ones relevant for flux compactifications). 
The AdS/CFT duality relates the two questions suggesting solving 
 them  together  rather than separately is the  best strategy. 
This is the modern   analog of the  ``harmonic oscillator'' or ``Ising''
problem  -- to solve 
two  most symmetric non-trivial 4-d gauge theory and  10-d superstring theory -- 
planar ${\cal N}=4$  SYM theory and  its dual --   free superstring  in \adss.
They  have  powerful  hidden symmetries 
allowing one  to solve  problem  ``in principle'' using integrability methods. 
 The  ${\cal N}=4$  SYM theory   has  
 maximal {supersymmetry}  and  {conformal  invariance} but 
it is a priori unclear in which sense it could  be integrable. 
Integrability might be 
 expected in the spectrum of anomalous dimensions
(as it was previously observed in YM 
 gluonic sector --  emergence of XXX spin chain  Hamiltonian 
 as 1-loop anomalous dimension operator \ci{lip}) 
 but  {hidden symmetries}  
should  play   broader role as they are 
``inherited''   via AdS/CFT  from 2-d integrable QFT --
 string $\sigma$-model. In that sense one  may 
 hope to  use 2-d integrable  QFT to solve 4-d CFT.  

{Superstring  sigma model  in \adss }
is  integrable in ``canonical'' sense being an example 
of a   {sigma model on symmetric space}.  Its 
classical equations  admit infinite number of conserved charges. In particular, it is 
closely related  (via Pohlmeyer reduction) to 
 (super) sine-Gordon  and non-abelian Toda models \ci{gt}. 
For example, special motions of strings  are described by 
  {{integrable}}  1-d  mechanical systems  (Neumann,  etc.).
The integrability of string theory  {extends} to {quantum  level}: 
we have  evidence of that   to 2 loops in $\a'$ expansion 
 using  AdS/CFT  ``bootstrap'' reasoning (see ref. \ci{rt2} 
 and refs. there).

{Quantum   integrability} of string theory sigma-model  {should  control}
the   spectrum of  string energies  on $R \times S^1$ 
(or anomalous  dimensions  of 2-d primary operators =
 vertex operators on $R^{1,1}$). It may also  have implications  for  
 correlation functions of  vertex operators  
(or closed-string scattering amplitudes) but  to which extent  
that might be true is unclear
(for example, in flat space the string action is just a collection of free 
oscillators but space-time amplitudes are rather nontrivial).

The quantum integrability of string theory for all values  of string tension ${{\sqrt \lambda} \over 2 \pi}$
 should then imply integrability 
on the gauge theory side for 
{any} value of 't Hooft  coupling 
$\lambda= g^2_{\rm YM}  N_c$. Again, this should be true for the spectrum of anomalous dimensions of single-trace  gauge invariant operators, while 
implications of integrability for correlation functions of such operators 
a priori seem  rather limited (though this may turn out 
 not be so given remarkable  hidden symmetries in the on-shell 
 gluon scattering amplitudes related to cusped Wilson loops discovered recently, see, e.g., \ci{korch}). 

Regarding  the spectrum of states, in the last 7 years  there
was an impressive  progress (for reviews, see, e.g. \ci{jp}).
%  based on superposing perturbative gauge and string theory results with assumption of exact  %integrability. 
The  {spectrum of ``long'' operators  or  ``semiclassical'' string states with large 
quantum numbers } is  now understood to be  described  
 by the {{ Asymptotic Bethe Ansatz }} (ABA) 
in its final BES \ci{bes}  form.
The ABA  was constructed using  information from perturbative gauge theory 
(spin chain for 1-loop anomalous  dimensions, ...)
and perturbative string theory (classical and 1-loop phase,...), 
 symmetries (magnon S-matrix), and the assumption of exact integrability.
 The  consequences of ABA were  {checked } against  all available perturbative  
 gauge and string theory data. 
The  key example is the ABA prediction for the 
{cusp anomalous dimension} or dimension $\Delta$ 
of   twist 2 operator $\Tr ( \Phi D^S_+  \Phi)$. Namely, 
 $ \Delta -2 \stackrel{S \to \infty}{=}    f(\lambda) \ln S  $ 
 where $f(\l)$ satisfies exact integral equation   solution of  which is known 
 in principle to any order in  small $\l$ or large $\l$ expansion 
 and agrees   with known perturbative results (see, e.g., \ci{kot,rt2,bkk}): 
\be && 
f(\lambda \ll 1) ={ \lambda \ov 2 \pi^2}\Big[ 1 - {\lambda \ov 48} + {11 \lambda^2  \ov 2^ 8\cdot  45}
- \big[ { 73 \ov 630} + { 4\z^2(3) \ov \pi^6}\big] 
{ \lambda^3\ov  2^ 7 } + ... \Big] \ , 
 \\  &&
f(\lambda \gg 1) 
={\sql\ov \pi} 
 \Big[1 - {3\ln 2  \ov  \sql}  - {K \ov  (\sql)^2}  - ...\Big]\ , 
\ee
where $K$ is Catalan's constant. 

 More recently, there was a substantial progress towards 
 understanding the  {spectrum of ``short'' operators, i.e.
 the spectrum  of  all quantum  string states.
 This  was achieved  solely on the  string side 
 by  using 2-d integrable  field theory methodology generalizing the 
 ABA  describing  states on $R^{1,1}$  
  to Thermodynamic  Bethe Ansatz (TBA)    which should be describing 
  states on $R \times S^1$  \ci{tba,gkv}. 
   It remains to be understood   why inclusion of wrapping diagram contributions
  to anomalous dimensions should lead to a similar modification from the gauge theory perspective. The construction of TBA is  rather non-trivial  due to 
  lack of 2-d Lorentz invariance 
in the standard ``BMN-vacuum-adapted'' l.c. gauge on string theory side. 
In few special  cases the  ABA  ``improved'' by Luscher corrections is enough:
4- and 5-loop Konishi operator dimension,\foot{The 5-loop result of \ci{janik}
was not so far reproduced  directly on the gauge theory side and  so remains a string-theory/integrability 
prediction.}
  and   4-loop minimal twist operator  dimension 
were computed \ci{janik}  in this way. 
What remains to do is to thoroughly check the TBA  predictions against 
perturbative string and gauge theory data. 

The  key example of a ``short'' operator is the Konishi operator \ci{kon} 
$\Tr ( \bar \Phi^i \Phi_i) $ 
for which the   weak-coupling anomalous dimension is  now known 
up to 5 loop orders\ci{zan,janik}: 
\be && \g(\lambda\ll 1 )
 = {12\l\ov (4\pi)^2} \Big[1   -  {4\lambda\ov (4\pi)^2}  +  {28 \lambda^2 \ov (4\pi)^4} \cr 
 && \ \ \ \ \ \ \ \ \ \ \ \ \ \ \ \ \ \ \ \ \ \
 \ \ \ \  -\ 
 [ 208 - 48 \zeta(3) + 120 \zeta(5)  ] {\lambda^3\ov (4\pi)^6} \cr
 && \ \ \ \ \ \ \ 
 +\ 
 8[ 158  + 72  \zeta(3) - 54 \zeta^2 (3) -90  \zeta(5)
     +  315 \zeta(7)   ] {\lambda^4\ov (4\pi)^8}
 + ... \Big]
%4-loop -- wrapping: 
\ee 
\def \bb {{\rm b}} 
The planar perturbation theory should have a finite  radius  of convergence;
suppose we  sum up this series   and re-expand at strong coupling $\lambda$ -- 
what should we expect to get? The AdS/CFT correspondence suggests
\ci{gkp1} that  $\g ( \lambda \gg 1) $ should start with $ \sim \fl  $ term. 
As was argued in \ci{rt} and will be discussed below, 
in general  string-theory arguments imply  that 
\be 
\g ( \lambda \gg 1) =  2 \fl  + %{\bb}_0 
b_0+  { b_1 \ov \fl  } +  { b_2 \ov (\fl )^2 }
+  { b_3 \ov (\fl )^3 } +... \ . \la{str} 
  \ee
 Here the 
values of b$_0,$ $b_1$ should be rational, while  $ b_3$ should be transcendental.
The analysis in  \ci{rt}  leads to 
\be 
%\bb_0 
b_0= -4, \ \ \   b_1 =1 \ , \ \ \ b_2=0, \ \ \ \la{res}
\ee
while $b_3$ should  contain $\zeta(3)$.  
At the same  time, the numerical result 
found from the  TBA/Y-system approach \ci{gkv} 
gives  $b_1$ that  is approximately twice as big. 
The reason for this disagreement remains to be  understood  but it is very encouraging that 
the numerical values  of $b_0, b_1$ in  \ci{gkv}   suggest that they are  integer, 
i.e. rational as it should be  according to  the string-theory logic. 

There   are many open questions   remaining:  
  (i) 
 which is radius of convergence of weak coupling expansion? \foot{The BMN dispersion relation 
 $e(p) = \sqrt{ 1 +  { \lambda \ov \pi^2} \sin^2 {p \ov 2} }$
  suggests an upper bound on the radius of convergence: 
  $ {|\lambda | \ov 4 \pi^2 }  <  { 1 \ov 4} $.}
 (ii) 
  how to find the analytic form of strong-coupling expansion from TBA/Y-system? 
 (iii) how to carry out   direct 
 matching of short operators onto  string spectrum found in near-flat-space expansion?
 (iv)  is there  ``level crossing''
  as one increases $\lambda$?
 (v) is strong-coupling expansion  of dimensions of short operators 
 Borel summable or not  as in the case of the cusp anomaly? 
 (vi) are there 
 exponential corrections  $ e^{ - a {\sql }}$  in  \rf{str}  
 like in the  cusp anomaly case \ci{bkk}?  
  There are also deeper  issues: 
 how to solve string theory from first principles --
 which are  fundamental variables? how to  preserve 2-d Lorentz  invariance?
 how to prove quantum integrability? 
 is there a useful 
 lattice version of string  ``supercoset''  sigma model?
 
 Below we shall concentrate on explaining the origin of 
 the expansion \rf{str} of the energy of a  string  state 
 which is dual to  Konishi operator following ref. \ci{rt}.

\section{Approaches to finding energies  of quantum strings in \adss}

According to AdS/CFT energies of quantum strings  in AdS 
should be the same as dimensions of the corresponding gauge theory operators. 
The aim  is to 
compute  the leading $\a' \sim  {1 \ov \sql} $  corrections to the 
energy  of  the 
 \ored{``lightest''   massive string state }
 which should be  dual to the 
\ored{Konishi operator} 
in SYM theory. 
 
%%%%%%%%%%%%%%%%%%%%%%%%%%%%%%%%%%%%%%%%%%%%%%%%%%%
%%%%%%%%%%%%%%%%%%%%%%%%%%%%%%%%%%%%%%%%%%%%%%%%
The members of the {Konishi multiplet}}
are operators related   to the lowest canonical dimension 
 %$SO(4) \times SO(6)$
  singlet  by supersymmetry. They  correspond to highest-weight states of
   $SO(2,4) \times SO(6)$ labelled  by (see, e.g.,   \ci{bia})
  $[J_2-J_3,J_1-J_2,J_2+J_3]^{\Delta_0}_{(s_L,s_R)}$ with the singlet being
  $[0,0,0]^2_{(0,0)}$. 
 Then full  dimension is 
 $\D= \D_0 + \g(\lambda)$,  \  $\D_0= 2,{5\ov 2},3,...,10$.
 All operators in supermultiplet 
have the  same anomalous dimension $\g$.\foot{That follows from 
the assumption of exact superconformal algebra in which the dilatation 
operator commutes with  supersymmetry generator $Q$ on $Q$ itself, i.e.
 $Q$ has exact conformal dimension 1/2. 
Then   members   of supermultiplet obtained from ``ground state'' 
by acting by $Q$ will have dimensions  differing by (half)integers.}
 Examples of such operators are 
 $\Tr ( \bar \Phi^i \Phi_i), \  (i=1,2,3)$ with  $\D_0= 2 $; \ \ \ 
 %+  \g\ , \ \  \  \g= O(\lambda)   $  
 $\Tr ([ \Phi_1,\Phi_2]^2)$   in $su(2)$  sector with  $\D_0= 4;\  $ %+  \g  $; \ \ \ 
 $\Tr (\Phi_1 D_{+}^2 \Phi_1)$  in $sl(2)$ sector   with $\D_0= 4 $.
  %+  \g  $, etc. 

\def \rR  {{\rm R}}

Assuming no ``level crossing'', the  Konishi operator
having lowest nontrivial dimension at weak coupling 
  should  be dual to 
the 
``lightest'' among  massive \adss   string states.
At
large $\sql = { \rR^2\ov \a'}$  a 
``small'' string at the ``center'' of  $AdS_5$ is in {nearly flat} space. 
At 
strong  coupling we are dealing with    perturbative  string theory: 
   string states are built out of ``flat-string''  oscillators
   and there is a 
large degeneracy of mass spectrum which is lifted once the curvature is switched on.  
{{In flat space case}} we have string masses   
  $ m^2 =  {4 ( n-1)\ov \a'} ,  \ \ \ 
   n= \ha ( N + \bar N) =1,2, ..., \ \   N= \bar N $.
The 
$n=1$ is    the massless  IIB supergravity (BPS)  level 
(l.c. vacuum $|0 >$:\    $(8+8)^2=256$ states). 
The {$n=2$} is the   first massive level; it  is 
highly degenerate: 
 $ [( a^i_{-1}  +  S^a_{-1}) |0 >]^2 = [( 8+8) \times (8+8) ]^2$.
Switching  on \adss  background fields lifts degeneracy:
states  with  ``lightest'' mass are thus at  \oblue{first excited  string level }
and thus 
should  correspond to  Konishi  multiplet.
%op. (and its susy descendents)

 The \ored{string spectrum in \adss} is organised \ci{bia}  in   
  long multiplets 
 of  $PSU(2,2|4)$:
 remarkably, flat-space string spectrum can be reorganised
  in multiplets of $ SO(2,4) \times SO(6) \subset PSU(2,2|4)$. Namely, 
   $SO(4) \times SO(5)  \subset SO(9) $ reps 
   can be 
    lifted to $SO(4) \times SO(6)$ reps  of $SO(2,4) \times SO(6)$, etc.  
   Then the 
   Konishi long multiplet  
   $\hat T_1 = (1 + Q + Q \wedge Q +...) [0,0,0]_{(0,0)}$
   determines the Kaluza-Klein  ``floor'' of 1-st excited string level: 
   $H_1= \sum_{J=0}^\infty [0,J,0]_{(0,0)}  \times  \hat T_1   $.
   States on the first excited level with $J\not=0$  are outside Konishi multiplet, i.e.
 they   should have higher anomalous dimension.

 For  scalar massive   state represented by a field  in $AdS_5$ one expects to find 
$(- \nabla^2 +   m^2) \Phi + ... =0$, i.e. 
$\D(\D-4) = (m \rR)^2 +  O(\a')= 4 ( n-1)  { \rR^2\ov \a'} + O(\a')  $
or 
$\D = 2 + \sqrt{ (m \rR)^2   + 4 + O(\a')} $ 
and thus \ci{gkp1}
  \be \ored{ \D(\lambda \gg 1)  =  \sqrt{ 4 ( n-1) \sql } + ...  } \ee 
Thus  for the first massive level 
 $n=2$: \ \ $\D = 2 \fl+ ... $.
What about 
subleading corrections?
In general, 
\oblue{comparison between  gauge and string theory states is non-trivial}.
In   gauge theory 
 for  \ ($\lambda \ll 1$) the     operators are built out of free fields,
 canonical dimension $\D_0$ determines states that can mix.
 In string theory with 
 ($\lambda \gg 1$) the   near-flat-space  string  states  are 
 built out of 
free oscillators,  and the level $n$   determines states that can mix.
One non-trivial  question is then about the 
meaning of $\D_0$ at strong coupling  and the 
meaning of $n$ at weak coupling.
 A possible strategy is to  relate  states with same  global charges and 
to  assume ``non-intersection principle'' \ci{pol}:
no level crossing for states  with same  quantum numbers 
as  $\lambda$  changes from strong to weak coupling. 

There are several 
{\ored{approaches to computation of corrections to string energies}}:
(i) \oblue{vertex operator approach}:
use \adss string  sigma model  perturbation theory 
to find 
leading terms 
in anomalous  dimension 
of  corresponding 
vertex  operator \ci{pol,tse03}; 
(ii) \oblue{space-time effective action approach}:
use near-flat-space expansion and NSR vertex operators 
to reconstruct 
 $\a'\sim {1 \ov \sql }$ corrections
 to   corresponding 
 massive string state equation  of motion \ci{bl}; 
(iii) \oblue{``light-cone'' quantization  approach}:
 start with  light-cone gauge  \adss\ string action 
  and 
compute corrections 
to energy of 
 corresponding flat-space 
oscillator string state \ci{mtt}; 
(iv)  \oblue{semiclassical approach}: 
identify short string  state as small-spin 
limit of 
semiclassical string state 
 -- reproduce 
 the structure of 
 strong-coupling corrections 
 to short operators \ci{tt,rt}.
 
%%%%%%%%%%%%%%%%%%%%%%%%%%%%%%%%%%%%%%%%%%%%%%%%%%%%%%%%%%%%%%%%%%%

\def \hg {\hat \g} 
%%%%%%%%%%%%%%%%%%%%%%%%%%%%%%%%%%%%%%%%%%%%%%%%%%%%%%%%%%%%%%%%%%%%%%%%%%%%%%%%%%%%%
%General curved background: 

To  find the spectrum  one should 
solve  the  marginality (1,1) conditions on vertex operators, i.e. to 
diagonalize the 2-d  anomalous dimension  operator.
For example,  the  scalar anomalous dimension  operator $\hg $
acts on   $T(x) = \sum c_{n...m } x^n... x^m $  or on the coefficients $c_{n...m }$.
$\hg $  is a differential operator in target space 
   % (cf.   gauge theory dilatation operator)
   found from the $\beta$-function for the  perturbation $T$
\be &&I= { 1 \ov 4 \pi \a'} \int d^2 z \ \Big[ 
% (G_{mn} + B_{mn})
G_{mn}(x) \del x^m \bd x^n 
   + T(x)   \Big] \cr
&&\beta_T = - 2T  -  {\textstyle{\a'\ov 2}}\ \hg\ T  + O(T^2)  \cr
&&
\hg =    \Omega^{mn}   D_m D_n   +...  + \Omega^{m...k} D_m   ...  D_k
 + ... 
   \cr
   &&\Omega^{mn}=   G^{mn}  
   %+ p_1  \a' R^{mn}   
%+    p_2 \a' H^m_{\ kl} H^{n kl}  
+  O(\a'^3) , \ \ \ \ \ \ \ \ \ \ \    \Omega^{....} \sim  \a'^n  R^p_{....}
\ee
Solving   $  - \hg\ T   +  m^2  T   =0$ is the same as 
  diagonalising   $\hg$.
Similar     approach   applies to 
 massless  (graviton, ...)   and  massive  states:  
e.g.,    $\beta^G_{mn} = \a' R_{mn} +  O(\a'^3)$   
gives  the Lichnerowitz operator as anomalous dimension operator 
\be
% &&R_{mn} ( G + h) = R_{mn} + \ha \hg^{kl}_{mn} h_{kl} +   O(h^2) \cr
%&&
(\hg h)_{mn} = - D^2 h_{mn}  + 2 R_{mknl} h^{kl}  - 2R_{k(m} h^k_{n)} + 
 O(\a'^3)  \ee
For  massive string states  in a curved background we should get  the action 
 \be && \int d^D x \sqrt{ g}\ \Big[  \P_{...}  
 (- D^2 +  m^2 + X ) \P_{...}  + ... \Big] \cr 
 &&
 m^2 ={\te{ 4\ov \a'}} (n-1) \ , \ \ \ \ \ \
  \ \   X = R_{....} + O(\a')   \ee
In the  case of  $AdS_5 \times S^5$  	background 													
%%%%%%%%%%%%%%%%%%%%%%%%%%%%%%%%%%%%%%%%%%%%%%%%%%%%%%%%%%%%%%
$$R_{mn}  - {\textstyle { 1 \ov 96}}    (F_5 F_5)_{mn} =0, \ \ \ \ \    R=0\ , \ \ \ \ \ \   F_{5}^2=0 $$
so for  a 10-d scalar field $\Phi$  the leading term in 
 $X$ should vanish.
 This apparently implies  that  the 
 leading $\a'$ correction to the  \oblue{scalar} string state  mass  should be zero: 
%should be zero
%then 
%for a scalar (singlet) state  we should have 
\be && \Big[- D^2  + m^2 + \os \Big] \P \ =0 \ ,   \ \ \ \ \ \ \ \qquad  \D= 2 + \sqrt{ 4(n-1)    + 4 +  \os  } \ , \cr
&&\D_{(n=2)}  = 2 +  2\fl\ \Big[1 + 
  \te{ 1 \ov 2\sql}  +  \oss \Big]\ .  \la{nnn}
   \ee
  There are possible subtleties in this argument, 
  so one should try to rederive this result 
  using other approaches. Also,  
  one should  compare this to predictions for 
 {\oblue{non-singlet}} Konishi descendant  states -- 
they should have  the same  anomalous dimension.
For example, 
 $\Tr [\P_1,\P_2]^2$    corresponding  to $SO(6)$  
(2,2,0)  state  $J_1=J_2=2$   should be described by a string state  with  a 
tensor wave function $\P_{mn;kl}$.  
 
\def \L {{ \rm R}}

%%%%%%%%%%%%%%%%%%%%%%%%%%%%%%%%%%%%%%%%%%%

{\ored{To find $\hg$  for tensor states  one may use effective action approach}}:
%%%%%%%%%%%%%%%%%%%%%%%%%%%%%%%%%%%%%%%%%%
derive equation of motion for a massive string field 
in curved  background 
from quadratic effective  action 
reconstructed from  flat-space  NSR S-matrix. 
%assume general covariance; 
%equivalence of  conformal invariance
 %conditions and equations of motion
An example is a   totally symmetric NS-NS  10-d tensor 
-- a  state on  the  leading Regge trajectory in flat space. 
It is described  by a
 symmetric tensor 
 $\Phi_{\m_1...\m_{2n}}$    in a metric+RR background \   ($m^2 = {4(n-1) \ov \a'}$) \foot{It is
 assumed here  that $\Phi_{\m_1...\m_{2n}}$ is traceless and $D^{\m_1} \Phi_{\m_1...\m_{2n}}=0$.
 In general, the quadratic terms may  also contain mixing of different types of fields 
 which we ignore here.}
\begin{eqnarray}
&&L=R-\ft1{2\cdot5!}F_{5}^2+ O(\a'^3)  \nonumber\\
&&\qquad  -\ft12( D_\mu\Phi D^\mu\Phi + m^2\Phi^2) +\sum_{k\ge1}
(\alpha')^{k-1}\Phi X_k (R,F_{5},D) \Phi+...
%\label{eq:schemlag}
\end{eqnarray}
The 
assumption is that   $ \a' n R \ll 1,$  i.e. $  n \ll   \sql $.
That corresponds to   a 
small massive string in the middle of $AdS_5$ so that   
near-flat-space expansion   should be applicable.
    The resulting  expression is \ci{bl} 
    ($a_i$ are from $AdS_5$ and $m_k$ are from $S^5$) 
\bea
 &&L  =
\ft12\Phi_{\m_1\cdots \m_{2n}} ( -  D^2  + m^2 )  \Phi^{\m_1\cdots \m_{2n}}
 \nonumber \\
&&\ \ \qquad   +\frac{n^2}{\L^2}\left(
\Phi_{a_1 a_2 \m_3 \cdots \m_{2n}}\Phi^{a_1a_2 \m_3\cdots \m_{2n}}
-\Phi_{m_1 m_2 \m_3 \cdots \m_{2n}}\Phi^{m_1m_2\m_3\cdots \m_{2n}}\right)
+... 
\eea
We   may  consider a particular tensor 
with $S$ indices in $AdS_5$ and $K$ indices in $S^5$ and then 
end up with the anomalous dimension  operator 
\be  &&
[ - D^2  + (m^2   +  {\te{ {K^2 -S^2 \ov  2 \L^2}}} )  ] \P =0 \ , \ \ \ \ \ \ \ \ 
D^2 = D^2_{AdS_5}  + D^2_{S_5}    \nonumber \\
&&      m^2 ={ \te {4\ov \a'} }( n-1) =  { \te {2\ov \a'}} ( S+K-2). 
%\ \ \  \ \ 2n = S +K  
   \ee 
This 
symmetric transverse traceless tensor  corresponds the  
highest-weight state with the 
  Young table labels  $(\D,S,0; K, J, 0)$ where $J$ is an additional KK momentum. 
   If we  
extract $AdS_5$ radius $\L$ and set $\sql = {\L^2 \ov \a'}$ then  \ci{bl}
\be && ( - D^2_{AdS_5}  +  M^2 ) \P=0\ ,  \cr
&& M^2 = 2 \sql (S+ K-2) + \ha ( K^2 -S^2)  + J(J+4) - K    \ee
Equivalently, the 
marginality condition for the corresponding vertex operator is 
\be && 0=  - \sql   (S+K-2)   \nonumber \\
&& \qquad  +   \ha \Big[  \D (\D-4)  + \ha S(S-2)  -     \ha K(K-2)      -   J (J+4)    \Big] + 
 \os  \ee
 The lowest 
\ored{BPS  level}  is \   $n = \ha (S + K)=1$
and the 
\ored{first massive level} is \  $n = \ha (S + K)=2$. 
 A
state from the   first massive  level on leading Regge trajectory
 is 
$S=K=2, \ J=0$  and then  
\be &&
\D=  2  +  \sqrt{ 4 \sql  +  4  +  \os }   
= 2 + 2\fl \Big[ 1 +      \te{ {1\ov 2\sql}} +   \oss \Big]  
 \ee
This is the same prediction as for a singlet scalar state
found above in  \rf{nnn}. 
The 
constant term  2  here corresponds to 
a  $\D_0=6$  operator ($\D_0- 4=2$)  \ci{rt}. 

\def \pa {\del}

\def \V {{\rm V}}

Let us  now review  how the above expressions may appear  directly in the  
%%%%%%%%%%%%%%%%%%%%%%%%%%%%%%%%%%%%%%%%%%%%%%%%%%%%%%%%%%%%%%%%%%%%%%%%
 {\ored{{vertex operator approach}}}.
%%%%%%%%%%%%%%%%%%%%%%%%%%%%%%%%%%%%%%%%%%%%%%%%%%%%%%%%%%%%%%%%%%%%%%%%%%
If we would like to    
 calculate  the 2d anomalous dimensions  from  ``first principles''  we should start
  with the \adss
superstring theory action \ci{mt}
% [Metsaev, AT 98]
\be && I= { \sql \ov 4 \pi}  \int d^2 \s  \Big[ -\del N_p \bd N^p +  \del n_k \bd n_k 
+ {\rm fermions\ } \Big]
\\
&& N_p N^p \equiv N_+N_- - N_u N_u^*  - N_v N_v^*  =1 \ , \ \ \ \ \   
 n_x n_x^*  + n_y n_y^* + n_z n_z^*=1 \cr 
&&
 N_\pm = N_0 \pm i N_5, \ \ \  N_u= N_1 + i N_2, \ ... , \ \ 
n_x=n_1 + i n_2, ... \no  \ee
and 
construct   marginal (1,1)   operators  in terms of $N_p$ and $n_k$.
For example, a    vertex operator for a dilaton field  (highest-weight  state) is 
\be \V_J   = 
   \  (N_+)^{-\D}\  (n_x)^J  \ \Big[  -\del N_p \bar \del N^p  + \del n_k  \bd n_k  + 
   {\rm fermions}\Big]
   \ee 
  where  
   $
   N_+ \equiv  N_0 + iN_5 = 
   { 1 \ov z} ( z^2 + x_m x_m) \sim  e^{i t}  $, \ 
 %\ , \ \ \ \   
 $
n_x \equiv  n_1 + i n_2 \sim  e^{i  \vp } $. The marginality condition gives 
%rotation 
   \be &&
   0= 2 - 2  + { 1 \ov 2\sql} [ \D(\D-4) - J(J+4) ]  + \oss\ , 
 \ee
i.e. $\D=  4 + J $ \  as it should be  for a BPS state. 
%should be no corrections to all orders --  BPS state
The   vertex operator  for  a bosonic  
 string state 
  on  the leading Regge trajectory 
%with spin  $S$   and energy $E$ 
 in flat space  is 
$$
\V_S = e^{ -i E t } \big( \pa \rx\bar{\pa} \rx \big)^{S/2}\   , \ \ \ 
\  \rx = x_1+ix_2  \  , \ \ \ \  \a'  E^2  = 2 (S-2)$$
Candidate operators for states on leading Regge trajectory
in \adss  are 
 \be   
 &&  \V_J =  (N_+)^{-\D}\big( \pa n_x \bar{\pa} n_x \big)^{J/2}\ , 
\ \ \ \ \ \ \ \ \     n_x \equiv  n_1 + i n_2  \ \\
&& 
\V_S(\xi) =  (N_+)^{-\D}\big( \pa N_u \bar{\pa} N_u \big)^{S/2}\ , 
\ \ \ \ \ \ \ \ \     N_u \equiv  N_1 + i N_2  \   \ee
where we ignore 
   fermionic terms 
and possible 
  $\a'\sim {1 \ov \sql} $ terms from  diagonalization of 2-d anomalous 
   dimension matrix. 
A non-trivial question is  
how such operators  mix with operators  with same  charges and  dimension.
In general,  $ \big( \pa n_x \bar{\pa} n_x \big)^{J/2}$ mixes  with singlets
   %($ p,q= 0,..., J/4$)
   $$ (n_x)^{2p + 2q  } (\del n_x)^{J/2 - 2p}   (\bar
\del n_x)^{J/2 - 2q }
( \del n_m \del n_m)^p ( \bd n_k \del n_k)^q $$
 For operators  like  
 $
O_{\el,s} =  f_{k_1...k_{\el} m_1...m_{2s}}  
n_{k_1} ...  n_{k_\el}  \d n_{m_1} \bd n_{m_2} ...
\d n_{m_{2s-1}} \bd n_{m_{2s}}
  $  this 
  question was studied, e.g., in ref.  \ci{weg}.\foot{  In the  
simplest case of the operator 
$ f_{k_1 ...k_\el} n_{k_1} ... n_{k_\el}$  with 
traceless $ f_{k_1 ...k_\el}$ 
it has  the 
 same anomalous  dimension   $\hg$ as its 
highest-weight
representative 
$
V_J = (n_x)^J $, i.e. 
$ \hg   %   2- { 1 \ov 2 \sql}     [ 5 J + J (J -1) ]  +...   
   =         2-    { 1 \ov 2 \sql}  J(J + 4)    + ... . 
$
This  is the same result as 
found for a  scalar  spherical harmonic that solves  the  Laplace  equation 
 on $S^5$.
Similarly  for the $AdS_5$ or $SO(2,4)$ model:
replacing $n^J_x$ and $\partial n_m \bd n_m $   with  
 $N^{-\D}_+$ and $\del  N^p \bd N_p $,  with 
   $J = - \D$ and  $  { 1 \ov \sql} \to - { 1 \ov \sql}$. 
For example, the 
dimension of \
 $N^{-\D}_+ \partial N^p \bd N_p $ is  
$\hg  =   { 1 \ov 2 \sql}    \D(\D - 4)  \ +    \oss$.}
An example of   higher-level scalar  operator is  
 $$  { N_+^{-\D} [(\d n_k \bd n_k)^r+ ... ]} \ , \ \ \ \ \ \ r=1,2,... $$
 for which \ci{krav}
 \be && 0= -2(r-1)   +  { 1 \ov 2\sql} [   \D (\D-4) +  2r(r-1)]  \cr
&& \ \ \ \   + 
{ 1 \ov (\sql)^2 } [{\te{ 2 \ov 3}} r (r-1) (r- {\te{7 \ov 2}})  + 4r ]  + ... \ . 
\ee
For 
   $r=2$ this is a candidate for a state on the  first excited  level.
   However,  the contribution of fermions may change  the value of the 2-d anomalous dimension.

   In general, the  2-d anomalous  dimensions are given by a {\it regular }
   series expansion in $\a' = { 1 \ov \sql}$, 
   while $1 \ov (\fl)^k$  appear as a result of solving 
   quadratic-type  equations for $E=\Delta$ following from the marginality condition.\foot{In particular,  there  cannot be any $\log \lambda$  terms that appear 
    in the strong-coupling expansion  of the anomalous dimensions 
   computed using  asymptotic Bethe ansatz equations \ci{brs}.}
   
As an  illustration, for  operators with two spins $J_1=K,\ J_2=J$ in $S^5$:
\be 
&& V_{K,J} =  N_+^{-\D} 
\sum_{u,v=0}^{K/2}  c_{uv}  M_{uv} \ ,   \cr
 && 
M_{uv } \equiv 
 n_y^{J-u-v} n_x^{u+v} (\partial n_y)^u (\partial n_x)^{K/2-u}   (\bd n_y)^v (\bd n_x)^{K/2-v  }   
 \ee  
the 
highest and lowest  eigenvalues of the  1-loop anomalous  dimension  matrix  are \ci{tse03} 
\be &&
%\hg_{\rm min} =
  2- K 
  +  { 1 \ov 2 \sql}  [   \D ( \D - 4)  -    \ha   K ( K + 10  )  -  J(J+4)  -  2  J K  ]  
  +  \oss\ ,   \cr
&&  
%\hg_{\rm max}  =  
  2- K 
  +  { 1 \ov 2 \sql}  [  \D ( \D -4)  -  \ha  K ( K + 6 )    -   J(J+4)   ] + \oss     \ee
  Again, the fermionic contributions 
  may  alter terms linear in $K$. 
 
 \oblue{The main question  on the way to finding energies or $\D$'s 
 of the corresponding states is thus  how  to take the \adss 
 fermionic contributions  into account? }
 % to make $K=2$ the zero  of $\g$   (BPS)  
To this end  we shall  follow ref. \ci{rt} 
and employ the semiclassical  approach based on the
 full superstring action.

%\renewcommand{\theequation}{3.\arabic{equation}}
% \setcounter{equation}{0}

%\setcounter{equation}{0} \setcounter{footnote}{0}

%%%%%%%%%%%%%%%%%%%%%%%%%%%%%%%%%%%%%%%%%%%%%%%%%%%%%%%%%%%%%%%%%%%%%%%%%%%%%%%%%%%%

{\section {{ Semiclassical approach: small-spin expansion}}}

In semiclassical expansion  for an energy of a  string state 
one fixes the 
``classical'' spin parameter 
$ \J = { J \ov \sql}$ and expands in  large string tension  $\sim \sql$, 
\be
 E= E( { J\ov \sql}, \sql)   =  \sql  \E_0 ( \J) +   \E_1 (\J)  + { 1\ov  \sql}   \E_2(\J) + ...  
\ee
In the ``short'' string limit   $ \J \ll 1$   
\be 
\E_n= \sqrt{  \J } \ ( a_{0n}   + a_{1n} \J + a_{2n} \J^2  + ... ) \
\ee 
Since   $\sql\gg 1 $ and $ \J= { J \ov \sql}$=fixed,  here 
  $ J \sim \sql  \gg 1$.
If we knew all the terms in this expansion  we  
could 
express  $\J$ 
in terms of $J$, fix  $J$  to a   finite value and re-expand  in $\sql$. Then 
we would get 
%[this is what we need to compare with  gauge theory BA results for short operators in strong coupling %expansion] 
%Rewriting the above   expansion  
%in terms of $J$  
\be 
E= \sqrt{ \sql J} \Big[ a_{00} + {  a_{10}  J + a_{01} \ov \sql }  + { a_{20} J^2  +  a_{11}   J + a_{02} \ov ( \sql )^2 } 
+ ...\Big] 
\ee
 To trust the coefficient   of $ 1\ov (\sql)^n$
 one would need to know   coefficients  of up to  $n$-loop  terms,  
e.g.,  the classical $a_{10}$  and the 1-loop   $a_{01}$   coefficients  are 
 sufficient in order  to fix the $ 1\ov \sql$ term.
 
 Since here we are  interested in a short string probing the near-flat-space limit, we may 
(i) start with  classical string solutions in flat space 
representing states  at 1-st excited string level, and then 
(ii) embed them  into \adss  and compute the 1-loop correction 
to  their energy. 
The two basic examples are \ci{rt}:
(1)  circular string with 2 spins in two orthogonal planes, and 
(2)  folded   string spinning in one plane. 

 \oblue{The rigid circular  string   rotating  in two  planes of flat    $ R^4$ }
 space is described by 
\bea 
&&t=\kappa\tau \ , \ \ \ \ \ 
{\rx_x}\equiv  x_1+ i x_2 =\  a \ e^{i ( \tau +\sigma) }\ , \ \ \ \ \ 
 {\rx_y}\equiv x_3+ i x_4 =\  a\  e^{i  ( \tau -\sigma) }  \ , \cr
&&
E_{\rm flat}  = {\te  {\kappa \ov \a' } } = \sqrt{ {\textstyle{4\ov  \a'}}  J }\ , \ \ \ \ \ \ \ \ 
 \ \  J_1=J_2=J =  {\te{  a^2 \ov \a'}} \ .   \eea
 Identifying  oscillator modes that are excited on this solution 
 we may 
  associate  it  with a 
  quantum string state  created by 
  %``target-space holomorphic'' 
 \be \la{vert} \ e^{-iEt}\ 
  \Big[(\del n_{x}  \bd \rx_{x})^{J_1\ov 2} (\del n_y  \bd \rx_y)^{J_2\ov 2} + ... \Big] 
   , \ \ \ 
   \a' E^2 = 2 ( J_1 + J_2 -2)    \ee 
  In the  $J_1=J_2$  case  to get the  quantum-state analog
  of classical expression one would need to  shift $J \to J-1$, i.e.  
 \be \la{fla} 
 E_{\rm flat}  =\sqrt{ {\textstyle{4\ov  \a'}}  (J-1) } \ . \no \ee
 Then $J_1=J_2=2$    corresponds to  state on 1-st 
  string level $n=2$. 
 
 A  folded string rotating in a plane in flat space is represented by 
\bea \la{ftf}
&& t=\kappa\tau \ , \ \ \ \ \ 
{\rx_1}\equiv  x_1+ i x_2 =\  a  \sin \s \   e^{i  \tau  }\ , \ \ \ \ \ 
% {\rx_2}\equiv x_3+ i x_4 =\  a\  e^{i  ( \tau -\sigma) }  \ , 
\cr
&&  E_{\rm flat} = \sqrt{ {\te{2\ov  \a'}}  S }\ , \ \ \ \ \ \ \ \ 
 \ \  S=  {\te{  a^2 \ov 2\a'}} \ .   \eea
 This is a 
semiclassical  counterpart of a  quantum  string state 
 on  the leading Regge trajectory  created by    the  vertex operator 
\be e^{-iEt}\ \Big[(\del \rx_x  \bar  \del \rx_x )^{S\ov 2} + ...\Big] \ , 
 \ \ \ \ \ \ \ \ \ \ 
\a' E^2 = 2 (S-2)  \ .  \ee
 There are 
3 obvious choices of  how to  embed the circular  solution  into 
 \adss:  
 (i) the  two 2-planes may  belong  to $S^5$:   
 $J_1=J_2$ ``small string''; 
 (ii) the two 2-planes  may belong  to $AdS_5$:  $S_1=S_2$ ``small string'';
 (iii)  one plane in  $AdS_5$
 and the other in  $S^5$:
     $S=J$ ``small string''.
There are similar  3 choices for the  folded string.
We may then study 
 each case in \adss, compute 1-loop correction to energy 
    and  interpolate to small values of $S,J$ and  try to   
 match to states in the  Konishi  multiplet table.
 Getting the same expressions up to a constant shift  would   
  verify the universality of the  strong-coupling expansion 
 of   the 4-d anomalous dimension   
 of  the dual gauge theory operators in the same supermultiplet.

 \oblue{The final result} found in \ci{rt} is 
 \be 
 && E\equiv \Delta =  2 \fl +  b_0  +  { b_1 \ov \fl}  + { b_2 \ov \sql} + { b_3 \ov (\fl)^3} ...
 \la{eee} \\  
 &&  b_0 = \D_0 -4 \ , \ \ \ \ \ \ \ \ \ \ \   b_1=1   \ee 
 where 
 $\D_0=4$ for the  3 circular string cases and 
 $\D_0=6$ for the  3 folded  string cases.
 The value of the coefficient 
 $b_2$ is sensitive to the 2-loop string corrections which were not computed 
 so far  but we 
 conjecture that that it is zero due to supersymmetry  (see ref. \ci{rt}).
% (it is hard to expect  vertex operator  approach). 

The simplest example is 
\oblue{ circular rotating string in $S^5$  with $J_1=J_2=J$} (see refs.  \ci{ft3,rt})
which  for    $J_1=J_2=2$  should be dual to the  Konishi descendant:  \ \ \ $\Tr ([\Phi_1,\Phi_2]^2) $.
\def \te {\textstyle}
The   above  flat  solution can be directly embedded into 
$R_t \times S^5$ inside of  $AdS_5 \times S^5$ -- it represents a 
string on a {\it small} sphere inside $S^5$ \ ($X^2_1 + ... + X^2_6 =1$)
% (e.g. $n=1$)
\be
&&X_1+iX_2=
%{\textstyle \frac{\sin\gamma_0}{\sqrt{2}}}\ 
a\ e^{i(\tau+\sigma)}, \ \ \ \   X_3+iX_4=
%{\textstyle{\frac{\sin\gamma_0}{\sqrt{2}}} }
a \  e^{i(\tau-\sigma)}~, \cr
&& X_5+iX_6=     \sqrt{1 - 2 a^2}          
%    \cos\gamma_0 \
, \ \ \ \ \ \ \  \ \ \ \ \ 
t=\kappa\tau~\ ,   
\cr
&&{\cal J}={\cal J}_1={\cal J}_2= a^2 
%{\textstyle \ha} \sin^2\gamma_0\
, \ \ \ \
~~
{\cal E}^2=\kappa^2=4{\cal J} \ . 
%E_0=\sqrt{\lambda}{\cal E}~,\ \ \ \ \  J=\sqrt{\lambda}{\cal J}\ ,
\ee
Remarkably, the exact classical energy   $E_0$ is just as in flat space 
\be 
E_0=\sql {\cal E}= \sqrt{4\sqrt{\lambda}J} \ , \ \ \ \ \ \ \ \ \ \ \ \ \ 
  J = \sql \J  
\ee
The 
{\oblue{1-loop quantum string  correction to the energy}}
is given by  the 
sum of the bosonic and  fermionic fluctuation  frequencies 
\be E_1=\frac{1}{2\kappa}\sum_{n=-\infty}^\infty \Big[ 4\omega_n{}+2n +  \omega_n{}_++\omega_n{}_- 
 -4(\omega_{n+}^f +\omega_{n-}^f)\Big] \ , 
\ee
where 
\be
&&AdS_5: \ \ \  \ \ \   \omega_n^2{}=n^2+4{\cal J} \no
\\
&&S^5: \ \ \ \ \ \  \omega_n^2{}_\pm = n^2 + 4(1 - {\cal J}) 
\pm 2\sqrt{4(1 - {\cal J})n^2 + 4{\cal J}^2} \no \\
&& {\rm Fermions}:   \ \ \ \ \ \   
 (\omega_n^f{}_\pm)^2  = n^2 + 1 + {\cal J} \pm \sqrt{4(1 - {\cal J})n^2 + 4{\cal J}}  \no \ee
Expanding  in small $\J$ and doing the  sums one finds  (UV divergences cancel) 
%normalize  to  flat space result in the  ${\cal J}\rightarrow 0$ limit:  
%in flat space  theory is gaussian -- trivial 1-loop correctio
\be
&&E_1
   =\frac{1}{\sqrt{\cal J}}\Big[ {\cal J}
    -{ 3 \ov 8} [1+8 \zeta(3)]{\cal J}^2 +\dots\Big]\cr
&&E=E_0+E_1= 2 \sqrt{\sqrt{\lambda} J}\Big[1+ \frac{1}{2\sqrt{\lambda}}
    -{\te{3 \ov 16}} [1+8 \zeta(3)] \frac{J}{(\sql)^2} + ... \Big]
  %  +\left(\frac{5}{2}+3\zeta(3)+15\zeta(5)\right)\frac{J}{\sqrt{\lambda}}+\dots
 %\right]\right]
\ee
To ensure the correct  short string limit  in flat space
we need to  shift $J \to J-1$.\foot{In general,  the correct  flat-space limit is 
consistent also with
$J \to J-1 + { a \ov \sql} + \oss$.}
In   $J=J_1=J_2=2$  case 
 that would  suggest that   for a Konishi state $[2,0,2]_{(0,0)}$
%($ 2J = J_1 + J_2 \to J_1 + J_2 -2 = 2 $) 
\be E = 2 \fl  \Big[1+  \te {\frac{1}{2\sqrt{\lambda}}}
   + \oss \Big]  \ee
The 
dual state in the Konishi table has $\D_0=4$. Thus  in \rf{eee} one gets 
     $b_1=1$ and     
 $b_2=0$ (at  1-loop  order) 
while 
$b_3$ is transcendental  (1-loop contribution to it contains $\zeta(3)$). 

The same  results
 are found for the  other (2  circular and 3  folded)  string solutions
 representing 5 other states at the 1-st massive string level 
 dual to 5 particular  operators in the Konishi multiplet table. 
We refer  to \ci{rt}  for the details.

\

%%%%%%%%%%%%%%%%%%%%%%%%%%%%%%%%%%%%%%%%%%%%%%%%%%%%%%%%%%%%%%%%%%%%%
{\section{ {  Conclusions}}}
%%%%%%%%%%%%%%%%%%%%%%%%%%%%%%%%%%%%%%%%%%%%%%%%%%%%%%%%%%%%%%%

Due to an  impressive recent progress  based on integrability methods 
we  are now at the beginning of understanding 
 the  quantum  string  spectrum in \adss 
  or the  spectrum of ``short'' operators in planar $\cal N$=4 SYM theory 
  for any value of `t Hooft coupling.   
 The  predictions of the integrability approach are still to be checked 
 against   perturbative string results. 
 This  requires better understanding of perturbative quantum \adss string 
 theory.
 % in \adss and, in particular,  its  near flat space expansion. 
 
 \

\section{Acknowledgments}
I  would like to thank R. Roiban for  a collaboration on ref. \ci{rt}
and comments on the draft of  this contribution.
I  also acknowledge useful  discussions with N. Gromov and R. Metsaev. 
I am grateful to the organizers of the Shifmania  workshop 
%"Crossing the boundaries: Gauge dynamics at strong coupling"
for their kind invitation and hospitality.

\bibliographystyle{ws-procs9x6}
\bibliography{ws-pro-sample}

\end{document}